**Thermodynamic Underpinnings Affecting Compressive and Tensile Plasticity in Refractory Compositionally Complex Alloys**

Eric A. Lass, Carla Joyce C. Nocheseda, Christopher Tandoc, Yong-Jie Hu, Peter K. Liaw

[1] Department of Materials Science & Engineering, The University of Tennessee Knoxville, Knoxville, TN 37996, USA

[2] Department of Materials Science & Engineering, Drexel University, Philadelphia, PA 19104, USA

* Corresponding Author: elass@utk.edu

**Abstract**

Refractory compositionally complex alloys (R-CCAs) show great promise as next generation structural materials for extreme conditions, but alloys with excellent high temperature mechanical properties most often exhibit virtually zero room temperature tensile ductility, hindering potential commercial application. Several semi-empirical metrics have been employed for predicting plasticity in R-CCAs, including valence electron concentration, Pugh ratio, Rice criteria, and the *D*-parameter. While many currently available ductility metrics correlate very well when compressive plasticity is considered, only the density functional theory (DFT) based *D*-parameter shows a strong correlation in tension. A new ductility metric is developed using CALculation of PHAse Diagrams (CALPHAD) thermodynamics, $\Gamma = (\delta_\chi \cdot \Delta G_{B\rightarrow H})/\gamma_{surf}$, where $\Delta G_{B\rightarrow H}$ is the free energy difference between the body-centered cubic (BCC) and hexagonal close-packed (HCP) phases, $\gamma_{surf}$ is the interfacial free energy of a (110) BCC surface, and $\delta_\chi$ is an "electronegativity-corrected" lattice distortion. The Γ parameter exhibits better correlation with tensile ductility than *D*-parameter and performs equally well in compression. While this improvement is modest, a thermodynamically grounded ductility metric is clearly required to capture intrinsic ductility behavior. Extrinsic contributions to brittle behavior in R-CCAs, namely interstitial oxygen content, are also briefly discussed. Combining the Γ ductility parameter with CALPHAD-calculated oxygen solubility is proposed as a first order attempt to capture both intrinsic and extrinsic contributions to ductile behavior in R-CCAs.

## 1. Introduction

Refractory compositionally complex alloys (R-CCAs), also often termed high-entropy or multiple principal-component alloys, have received significant research attention over the last 15+ years due in large part to their potential as next-generation structural materials for extreme environments beyond Ni-based superalloys, the gold standard for high temperature, extreme environment materials for over half a century. Single-phase body-centered cubic (BCC) alloys are the most studied [1–9], although multiphase alloys have also demonstrated promise for high temperature applications, most notably two-phase BCC plus ordered B2 phase [10–17]. Among the issues that must be addressed if commercialization of R-CCAs is to be successful is developing materials that combine excellent high temperature mechanical properties with reasonable room temperature ductility and fracture toughness [1–5,18,19]. In terms of single-phase BCC alloys, research on R-CCAs to date has identified materials that generally fall into two categories: 1) "soft" R-CCAs, or those with reasonable to good room temperature ductility but that exhibit a dramatic drop off in strength at elevated temperatures, usually around 1000 °C [1,9,19–26]; and 2) "hard" BCC R-CCAs, materials that possess good strength retention at elevated temperatures but exhibit virtually no ductility at room temperature [1,6,27–30]. The former class consists largely of derivatives of the so-called Senkov alloy, equiatomic HfNbTaTiZr, and contain large amounts of elements in Group IV of the periodic table, i.e., Hf, Ti, and Zr [20]. The latter class of alloys contain high fractions of Mo and W, along with Nb and Ta, with considerably less, if any, Hf, Ti, and Zr [1].

The difference in mechanical behavior between hard and soft R-CCAs lies largely in their dislocation core structure: soft alloys have a spread-out core while hard alloys have a tight, compact core like those found in Mo and W [8,9,22,24,28,31–40]. While a detailed discussion of the differences between hard and soft R-CCAs is outside the scope of this study, changes in dislocation core structure, and subsequently dislocation mobility, are related to *d*-band electron filling and changes in bond characteristics as the number of *d*-electrons increases [24,41]. Group VI elements (e.g. Mo and W) have five unpaired electrons in the *d*-band subshells, leading to more rigid, directional-type bonding, evidenced by their high elastic constants, which also helps to stabilize the non-close-packed BCC crystal structure. On the other hand, decreasing the number of *d*-electrons, moving to Group V and then to Group IV elements, results in less rigid bonds with a much higher metallic character, the reason Group IV elements exhibit a hexagonal close-packed (HCP) crystal structure in their standard state.

Several criteria for predicting ductility in R-CCAs have been proposed [25,41–51]. The valence electron concentration (VEC) is the easiest and most often used metric, where a VEC of $<4.5$ is expected to be ductile [25]. The Pugh ratio, $G/K$ (sometimes $K/G$), uses the shear and bulk moduli, $G$ and $K$, respectively, as surrogate parameters for plastic deformation and brittle failure; a smaller value indicates an increased likelihood for plastic deformation (i.e., ductility) [46]. Rice proposed that ductility is governed by the competition between dislocation nucleation from a crack tip and cleavage, giving rise to the Rice criteria, $Gb/\gamma_{surf}$, a smaller number is indicative of ductility [47,48]. Pettifor proposed the Cauchy pressure, the difference between elastic constants, $c_{12} - c_{44}$, as a ductility metric; a positive value indicates nondirectional metallic bonding and intrinsic ductility [43–45]. More recently, Hu et al. utilized density functional theory (DFT) to build on the work of Rice and proposed the "$D$-parameter", $D = \gamma_{surf}/\gamma_{usf}$, where $\gamma_{surf}$ and $\gamma_{usf}$ are the surface and unstable stacking fault energies, respectively [49,50]. Singh et al. recently developed a criterion based on local lattice distortion (LLD), or average atomic displacements, determined from density functional theory (DFT) calculations [22,51,52]. This method is similar to the atomic size mismatch parameter, $\delta$, in Eq. (1) below. However, LLD accounts for differences in elastic moduli, electronegativity between elements, and changes in atomic behavior with local environment to provide a physics-informed measure of lattice distortion.

The purpose of the present study is to propose and develop a thermodynamic basis for the observed mechanical properties of R-CCAs, distinguish differences between the two R-CCA types, and identify new ways of estimating alloy ductility *a priori*. This is accomplished by first examining the shortcomings of established criteria in predicting ductility and presenting observed correlations between reported compressive and tensile plasticity and thermodynamic properties. Second, the well-known relationship between stacking-fault energy (SFE) and deformation behavior in face-centered cubic (FCC) metals [53,54], for instance Fe-Mn-steels, is briefly introduced, noting that SFE is directly connected to the thermodynamic Gibbs free energy difference between FCC and HCP phases; a concept that has been utilized to successfully predict martensite transformation temperatures in FCC materials for not only HCP martensite, but also BCC/BCT (body-centered tetragonal) martensite, particularly in steels [55,56]. This concept is extended to BCC alloys, where partial dislocations are not observed, but where several displacive transformations between BCC and other phases can be described in terms of the splitting of full BCC dislocations over multiple neighboring planes, creating regions of local atomic stacking similar to HCP, face-centered orthorhombic (α" martensite or O'), or ω phase [57]. Such local structures have been reported in molecular dynamics simulations of R-CCAs [33,58]. The correlations between several thermodynamic and physical properties are examined, and a new ductility metric, founded in thermodynamics, is proposed. It should be noted that it is *not* the intent of the present work to predict mechanical properties, although the vision is that the theory put forth leads to advances in quantitative prediction of strength, plasticity, and high temperature mechanical behavior of R-CCAs.

## 2. Methods

### *2.1. Calculation of materials properties and existing ductility metrics*

The ductility metrics Pugh ratio (*G*/*K*), Cauchy pressure ($c_{12} - c_{44}$), and Rice criteria ($Gb/\gamma_{surf}$) were all calculated using a molar (RoM) value of the constituent materials properties. All pure element properties were taken from Springer Handbook of Materials Data [59], except Pauling and Allen electronegativity, which were taken from WebElements.com [60] and Wikipedia [61], respectively. *It should be noted that both the bulk and single crystal elastic properties for Group IV elements (Hf, Ti, an Zr) are properties of these elements in the HCP phase and do not necessarily reflect BCC values.* Despite use of HCP properties of Group IV elements being the convention in much of the literature on CCAs, it introduces uncertainty in the calculated materials properties for alloys containing Hf, Ti, and Zr and is one drawback of utilizing a RoM approximation instead of employing computational tools such as DFT. However, no reliable DFT-based estimates for elastic properties of pure Group IV elements in the BCC phase were found in literature. Further, if such values exist, a full set of DFT-calculated elastic properties for all refractory elements from a single source would be desirable to ensure consistency of properties between the elements. Therefore, HCP elastic properties are utilized here for Group IV elements in line with previous literature and to maintain a consistent source of elastic properties across all elements considered. All pure element materials properties used in the present work are available in Supplemental Materials (SM) Table S1.

The surface energy for the Rice parameter was estimated as $\gamma_{surf} = 0.15\,\Delta H_a/\rho_{110}$ [62], where $\rho_{110}$ is the molar density of (110) planes (i.e., the closest-packed, lowest energy planes of BCC), and $\Delta H_a = \Delta H_f + \Delta H_v$, the molar enthalpy of sublimation/atomization, where $\Delta H_f$ and $\Delta H_v$ are the enthalpies of fusion and vaporization, respectively. $\Delta H_f$ and $\Delta H_v$ where again taken from the data tabulated at WebElements.com. The average atomic diameter, $\bar{d}$, was calculated as a RoM of the atomic diameters of constituent elements, $d_i$, from which $\rho_{110}$ and the Burgers vector, *b*, were determined. The shortest interatomic distances in BCC Hf, Ti, and Zr tabulated in the Springer Handbook of Materials Data were used for $d_i$ for these elements instead of the interatomic spacing determined from their HCP lattice parameters to best represent the elements diameters in BCC solid solutions. For completeness, the calculations were also performed using the interatomic spacing determined from the HCP lattice parameters for Group IV elements. This resulted in only moderate changes to resulting correlation coefficients between ductility metrics (*b* in Rice parameter) and strain to failure and are available in SM Table S2, along with other materials properties calculated using HCP lattice parameters (i.e., atomic diameters and lattice distortions).

In addition to elastic and thermodynamic properties, other materials properties, e.g., electronegativity, VEC, etc., were calculated using the same molar RoM described above. The "HCP fraction" is simply the mole fraction of Group IV elements in the alloy. The lattice distortion, δ, given by

$$\delta = \left[ \sum_{i}^{n} c_i \left( 1 - \frac{d_i}{\bar{d}} \right)^2 \right]^{\frac{1}{2}} \quad (1)$$

where $c_i$, $d_i$, and $\bar{d}$ ($= \sum_i c_i d_i$) are the concentration, atomic diameters of the individual components, and average atomic diameter, respectively, were calculated using the BCC $d_i$ for all elements as described above.

To try and capture the role of local chemical environment on atomic diameter, similar to LLD, an electronegativity-corrected lattice distortion, $\delta_\chi$, is also defined. It is calculated in the same manner as δ in Eq. (1) except that $d_i$ is replaced by an "electronegativity-corrected atomic diameter" $d_{\chi i}$, given by

$$d_{\chi i} = d_i \left( \frac{\chi_i}{\bar{\chi}} \right) \quad (2)$$

where $\chi_i$ is the electronegativity of component *i* and $\bar{\chi}$ is the RoM electronegativity of the alloy. The reasoning behind $d_{\chi i}$ is: if the electronegativity of a component of the alloy is lower than the electronegativity of the rest of the components in the alloy, the electron density around the nucleus will be lower than the component by itself, leading to a net positive 'ion' with a smaller size than the neutral atom, and vice-versa. While it is not as rigorous as the DFT-calculated LLD, $\delta_\chi$ accounts for changes in atomic size and lattice distortion due to differences in local environment in an alloy compared to a pure element.

All ductility metrics, thermodynamic, elastic, atomic, electronic, and derived elastic quantities calculated for the alloys investigated in this study are provided in SM Tables S3-S14.

### 2.2. *Gibbs free energy and mechanical behavior*

#### 2.2.1. Rationale

Akin to the relationship between stacking fault energy (SFE) and the Gibbs free energy difference between FCC and HCP phases in FCC alloys [53–56], e.g., Fe-Mn steels, the Gibbs free energy difference between BCC and HCP phases, $\Delta G_{BCC\rightarrow HCP}$ ($\Delta G_{B\rightarrow H}$), is used in the present work as a proxy for deformation mechanism and dislocation mobility. The hypothesis is that this free energy difference is related to the "ruggedness" of the energy landscape through which dislocations must navigate during plastic deformation. This relationship is relatively easy to visualize in the case of FCC metals and $\Delta G_{FCC\rightarrow HCP}$ because there is a clear relationship between the two phases by the stacking sequence of close-packed planes, *ABCABC* for FCC and *ABABAB* for HCP. Slip of a perfect $\frac{a}{2}\langle 110\rangle\{111\}$ FCC dislocation requires adjacent planes to temporarily pass through a local configuration of HCP crystal structure, which manifests as a stacking fault bounded by two Shockley partial dislocations with $b = \frac{a}{6}\langle 211\rangle$. The energy distance between Shockley partials is directly related to $\Delta G_{FCC\rightarrow HCP}$ and is connected to the mobility of dislocations and the ease of slip and plastic deformation in FCC metals.

The relationship between thermodynamic phase stability and plastic deformation is less obvious in BCC materials where the phase relationship with other crystal structures is less apparent. However, if one considers the pathway of an $\frac{a}{2}\langle 111\rangle$ BCC dislocation during slip, it becomes apparent that the local atomic structure between adjacent atomic planes passes through configurations similar to other crystal structures [55,57]. If for instance, an $\frac{a}{2}[\bar{1}11](110)$ BCC dislocation dissociates into three "fictitious" partial dislocations, one $\frac{a}{4}[\bar{1}12](110)$ partial and two $\frac{a}{8}[\bar{1}10](110)$, as shown in Figure 1, an HCP-like stacking can be achieved by a shear of $\frac{a}{8}[\bar{1}10](110)$ on one plane followed by a shear of $\frac{a}{4}[\bar{1}12](110) + \frac{a}{8}[\bar{1}10](110)$ on an adjacent plane. The partial dislocations are fictitious because they are not observed in real systems. The stacking is HCP-like because atoms in one plane sit in the well between a triangle of three atoms in the plane below like in HCP, except that the triangle of atoms is not close-packed and equilateral. Note that the end result of this series of shears is the same as the $\frac{a}{8}[\bar{1}10](110)$ shuffle in the shear-shuffle model of the BCC-to-HCP martensitic transformation [57], i.e., the Burgers transformation pathway between BCC and HCP structures. The shear portion of this model occurs by an $\frac{a}{2}[\bar{1}1\bar{1}](\bar{1}12)$ splitting into six $\frac{a}{12}[\bar{1}1\bar{1}](\bar{1}12)$ on consecutive $(\bar{1}12)$ planes, turning the non-close-packed HCP-like structure into a true HCP crystal structure (Figure 1).

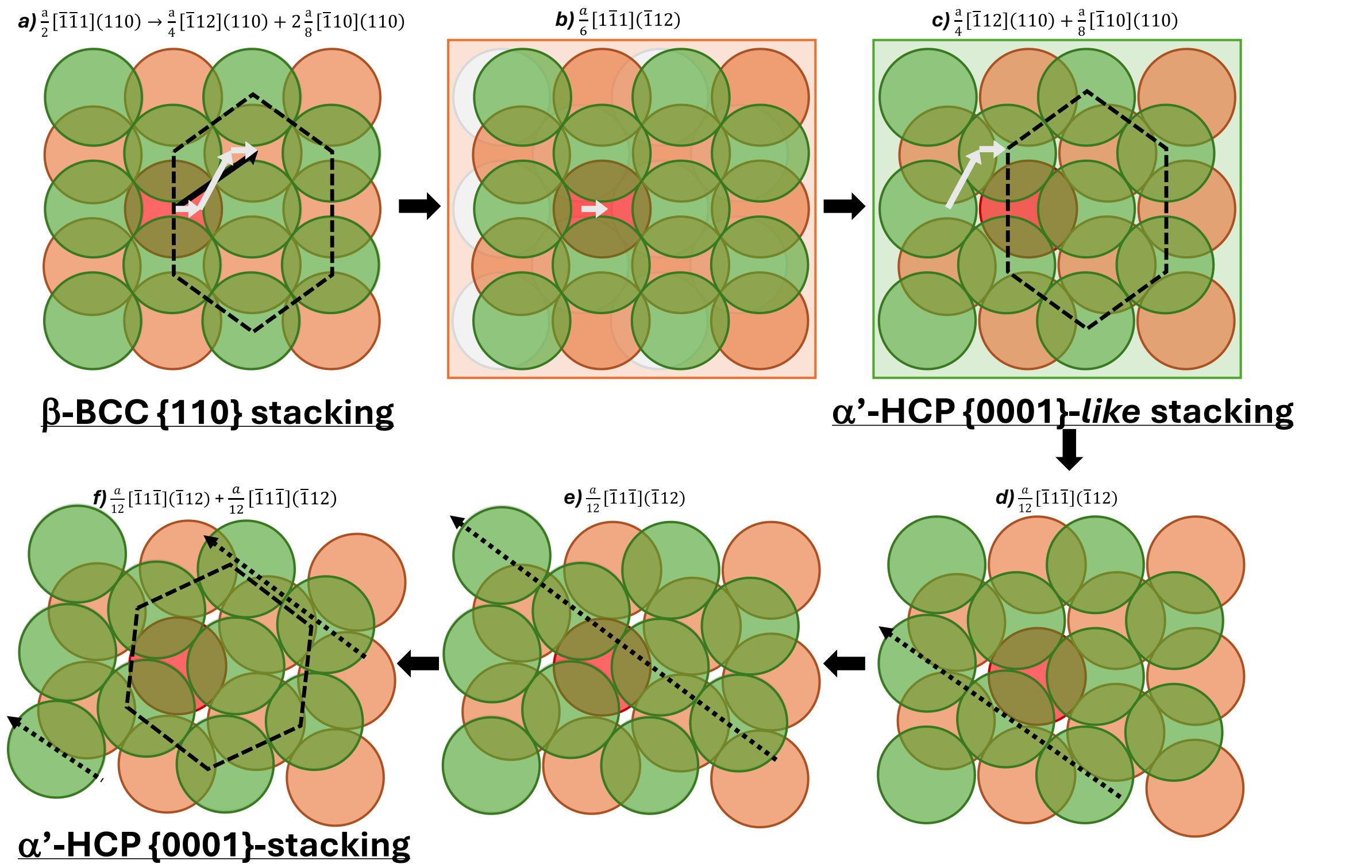

*Figure 1: Formation of HCP {0001}-like stacking by the dissociation of a BCC* $\frac{a}{2}[\bar{1}\bar{1}1](110)$ *into one* $\frac{a}{4}[\bar{1}12](110)$ *partial and two* $\frac{a}{8}[\bar{1}10](110)$ *partials, shearing first the orange and green planes by* $\frac{a}{8}[\bar{1}10](110)$*, then the green plane by* $\frac{a}{4}[\bar{1}12](110) + \frac{a}{8}[\bar{1}10](110)$ *(top); subsequent formation of HCP from the intermediate stage though the* $a/12\,[\bar{1}1\bar{1}](\bar{1}12)$ *shearing of adjacent* $(\bar{1}12)$ *planes.*

If the second described shear above occurs without the shear/shuffle of $(110)$ planes the orthorhombic α" phase is formed as shown in Figure 2, which is known to form martensitically in BCC-stabilized Ti- and Zr-alloys and in R-CCAs. Also shown in Figure 2, if a dislocation of the opposite sign on the same $(\bar{1}12)$ plane separates into three equal partials, i.e., $\frac{a}{2}[1\bar{1}1](\bar{1}12) \rightarrow 3\frac{a}{6}[1\bar{1}1](\bar{1}12)$ spaced on adjacent $(\bar{1}12)$ planes a $\{112\}$ BCC deformation twin is the result. Similar arguments can be made regarding the relationship between "partial" BCC dislocations and other phases, such as ω and O' While these partial dislocations may not exist in real systems, the fact that they form local atomic arrangements similar to other crystal structures and are consistent with the experimentally observed shears during displacive transformations and twinning indicates that a relationship between the free energy difference between BCC and these other phases may be related to dislocation mobility, analogous to the known relationship in FCC materials. Indeed, this relationship has been exploited in computational thermodynamic modeling to successfully describe the dislocation mediated martensitic transformation from FCC to BCC/BCT and BCC/HCP martensite in steels and to understand deformation in austenitic FeMnAl and Hadfield steels, even when a deformation-induced martensitic transformation is not observed [56,63].

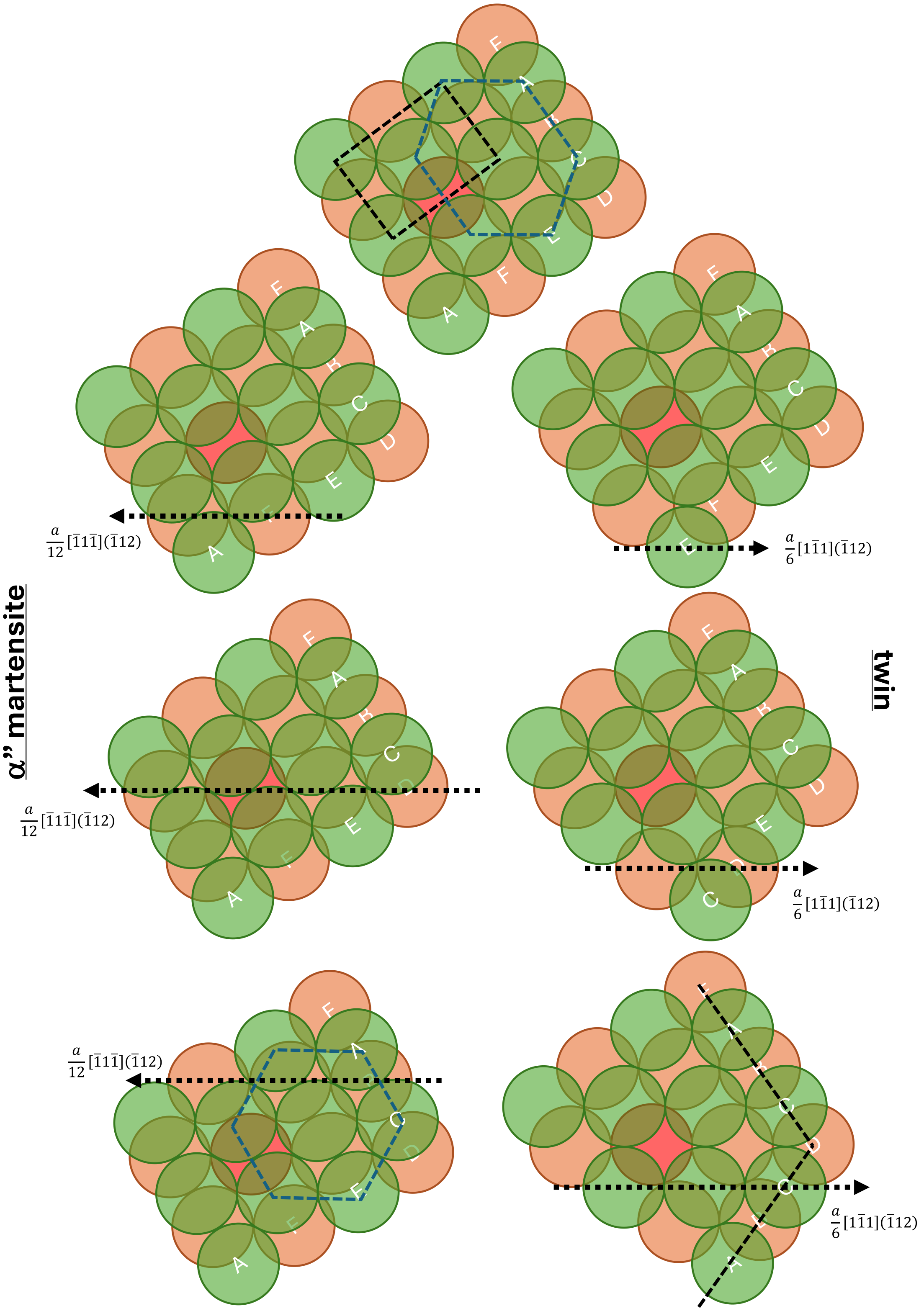

*Figure 2: Transformation pathways, viewed perpendicular to a {110} plane, from BCC to α'' martensite (left) and from BCC to a twin (right) by way of splitting the shear displacement of a full* $\frac{a}{2}[1\bar{1}1](\bar{1}12)$ *into: 1) three* $\frac{a}{6}[1\bar{1}1](\bar{1}12)$ *displacement on adjacent* $(\bar{1}12)$ *planes (twin) or 2) six* $\frac{a}{12}[\bar{1}1\bar{1}](\bar{1}12)$ *on every other* $(\bar{1}12)$ *plane (α'') for the α'' martensite and twin, respectively.*

### 2.2.2. Calculation of $\Delta G_{BCC \to HCP}$

Two CALPAHD-based computational software packages and databases were utilized to calculate the Gibbs free energies of BCC and HCP phases for each alloy composition: 1) the Thermo-Calc software and TCHEA8 database [63,64] and 2) the Pandat software and PanRHEA2024 database [65,66]. The TCHEA8 and PanHEA2024 databases do not contain thermodynamic descriptions for secondary solid solution phases, such as orthorhombic α” (this phase is said to be combined with HCP in the TCHEA8 database information sheet) and ω, so only BCC and HCP were considered. Calculations were performed at 373 K, a compromise to balance the desire to quantify $\Delta G_{B \to H}$ at the deformation temperature of interest, i.e., room temperature, and the potential loss of accuracy when extrapolating CALPHAD to temperatures where no experimental data exist. Each phase was calculated separately without using global minimization to ensure no phase separation occurred due to the BCC miscibility gap, and the free energy of the exact alloy composition was calculated for each phase. The disordered BCC_A2 phase was also used to avoid any additional contribution due to B2 ordering, real or a fictitious result of combining binary/ternary thermodynamic descriptions into a multicomponent database. The calculated BCC and HCP free energies for the alloys considered in the present work are provided in SM Tables S15-S17.

### *2.3. Calculation of D parameter*

The *D*-parameters of the alloy compositions investigated in this work were calculated using a recently developed machine learning (ML) model [49,50]. This model employs a set of physics-informed descriptors derived from DFT calculations, specifically designed to capture the local atomic bonding environments in randomly mixed BCC solid-solution alloys. Trained on DFT-calculated surface energies ($\gamma_{surf}$) and unstable stacking fault energies ($\gamma_{usf}$) of binary and ternary alloys, the ML model has demonstrated its ability to accurately predict these defect energies for multicomponent alloys beyond the training dataset within a 10-component chemical space. In the present work, the ML model was employed to predict $\gamma_{surf}$ and $\gamma_{usf}$ for alloy compositions with available experimental tensile data. The *D*-parameter was subsequently calculated as the ratio of $\gamma_{surf}$ to $\gamma_{usf}$ and used to assess the intrinsic ductility of these alloys. Calculated *D*-parameters for the alloy compositions investigated in the present work are included in SI Table S3-S5.

### *2.4. Experimental procedure*

While not the focus of the present work, experimental investigation of several R-CCAs designed using the theories developed herein occurred concurrently. These alloys were fabricated by vacuum arc melting the appropriate amount of the pure elements (>99 % purity) under an argon atmosphere. A Zr getter was used during each melting step to scavenge oxygen remaining in the chamber. Each sample was remelted a minimum of five times to ensure complete mixing. On the final melting procedure, most alloys were cast into permanent molds; samples containing W could not be readily cast and were investigated in the as-solidified button ingot. Some of the alloys were thermomechanically processed (TMP) via cold rolling and annealing. Tensile stress strain behavior was evaluated on both TMP and as-cast samples. Details of the processing and characterization of these materials will be included in future publications. A list of these alloy compositions and the processing that was performed on each are available in SM Table S18. These alloys were identified and experimentally characterized parallel to the present work and were not used in the analysis of the effectiveness of any ductility metrics or materials properties.

### *2.5. Statistical Analysis*

The correlation between compressive and tensile failure strain and the various ductility metrics and other materials properties were quantified using Pearson correlation coefficients. The Pearson correlation coefficient, denoted as $R_P$ here to indicate its relationship to the common coefficient of determination $R^2$ used to determine the goodness-of-fit to a linear regression, is a measure of the linear correlation between two parameters. A value of +1 indicates a perfect linear correlation, while a value of -1 indicates a perfect inverse linear correlation. While the interpretation of a correlation coefficient is subjective and depends on the nature of the data, an $|R_P| > 0.5$ is often considered a good correlation. For the purposes of the present analysis, we will rate the correlation of two variables as follows: *i*) excellent or very strong for $|R_P| > 0.8$, *ii*) very good or strong for $0.6 < |R_P| < 0.8$, *iii*) good or reasonable for $0.4 < |R_P| < 0.6$, *iv*) weak for $0.2 < |R_P| < 0.4$, and *v*) very weak to no correlation for $|R_P| < 0.2$. The calculated correlation coefficients between compressive and tensile plasticity and plasticity metrics and other calculated materials properties are provided in SI Table S19-S23.

It should be emphasized here that the condition of the material, i.e., fabrication and processing route, have a considerable effect on the observed mechanical properties of alloys. This is because microstructure characteristics are intimately dependent on processing, such as compositional inhomogeneity (e.g., solidification segregation), grain size, and particularly interstitial content (i.e., C, O, and N) in the case of R-CCAs. It is inevitable that alloys produced in nominally identical manners but by different research groups, or even different researchers within the same research group, will have some differences in processing conditions that may affect measured mechanical performance. Details such as grain size or interstitial content are often not reported in literature, making it impossible to track the effects of such variables in such a diverse dataset. Therefore, fabrication and processing details have been ignored for simplicity in the current study. However, the role of interstitial elements is discussed. It should also be noted that only total elongation to failure is considered in the present work. In cases where total elongation was not reported, or where plastic strain and/or uniform elongation reported, the largest of the reported strains is used.

## 3. Results

### *3.1. Ductility Correlations with Existing Metrics*

Figure 3 presents experimentally measured compressive and tensile failure strain as a function of the various ductility metrics described above. Failure strain data was collected from that tabulated by Borg et al. [18], Senkov et al. [19], and Singh et al. [51], and are listed in SI Table S2. Also shown are several compositions produced by the present authors. Pugh ratio and Rice parameter both show excellent inverse correlation with compressive failure strain, both with $R_P < -0.8$. VEC and $D$ exhibit very good correlations with $R_P$ of -0.68 and +0.71, respectively, while the fraction of HCP elements shows a reasonable correlation with compressive failure strain. Notably, there is no real correlation between compressive failure strain and Cauchy pressure, $R_P \approx -0.03$. Further, all values of the Cauchy pressure are positive, as expected for metals, while the Cauchy criterion for brittle behavior is $(c_{12} - c_{44}) < 0$. Clearly Pettifor's Cauchy criterion is not a valid indicator of ductility in R-CCAs or metals in general.

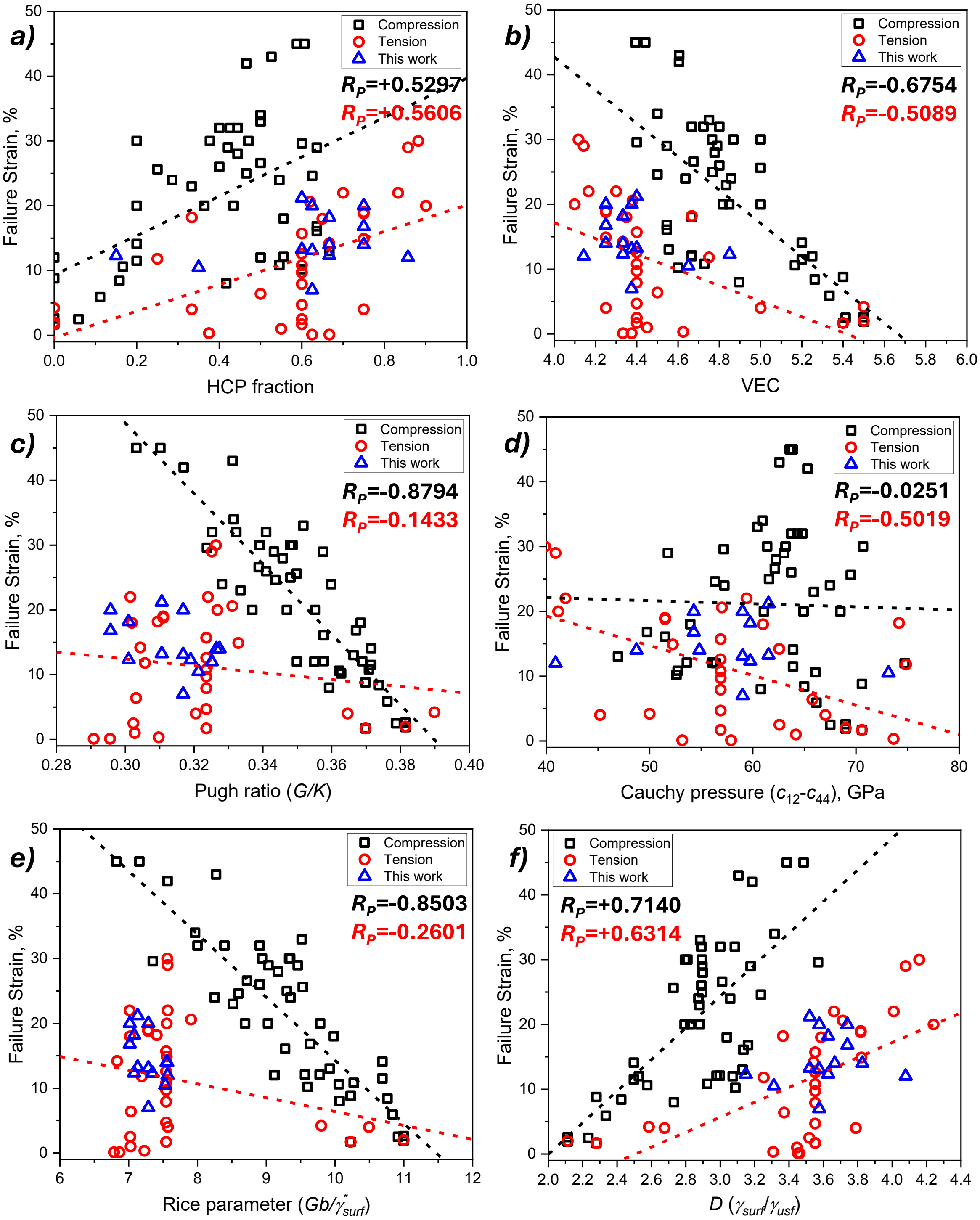

*Figure 3: Linear fit between experimentally measured compressive (black squares) and tensile (red circles) strain to failure and a) fraction of HCP (Group IV) elements, b) VEC, c) Pugh ratio (G/K), d) Cauchy pressure, ($c_{12}$-$c_{44}$), and e) Rice parameter ($Gb/\gamma^*_{surf}$), and f) D-parameter ($\gamma_{surf}/\gamma_{usf}$). Note $\gamma^*_{surf}$ was estimated using the enthalpy of sublimation and average atomic diameter in the Rice parameter while $\gamma_{surf}$ was calculated using DFT for D-parameter.*

The correlations between the various ductility metrics and *tensile* plasticity are considerably worse than those found in compression. *D*-parameter shows the strongest correlation with tensile ductility among the existing ductility metrics with an $R_P$ of +0.63. The Pugh ratio and Rice parameters exhibit weak to very weak correlation with tensile ductility, with $R_P$ of only -0.15 and -0.26, respectively. VEC and Cauchy pressure both exhibit good correlation with tensile ductility, $R_P \approx -0.5$. While the fact that VEC performs fairly well in both compression and tension is expected, the good correlation of Cauchy pressure with tensile failure strain is puzzling, considering it exhibits virtually no correlation in compression. Somewhat surprisingly, the simple metric of HCP mole fraction exhibits a stronger correlation with tensile ductility than any of the other metrics except *D*, with $|R_P| = 0.56$. This is not entirely unexpected since it is very closely related to VEC, and nearly identical for alloys comprised of only Group IV and V elements, which is true for nearly all reports of tensile ductility in literature. Indeed, several publications have discussed the strong correlation between tensile ductility and the fraction of Group IV elements [9,23,33,58,67]. Not shown in Figure 3 is LLD. While its correlation with tensile ductility is remarkably strong, $R_P = +0.88$ (SM Table S19), only six data points are available.

### *3.2. Ductility Correlations with Materials Properties*

Each of the ductility metrics presented in Figure 3 are similar in that they utilize one or more materials properties, often a pair of parameters meant as proxies for brittle and ductile fracture. As pointed out above, Pugh ratio uses *G* as a measure of ductile failure and *K* of brittle failure, for Cauchy pressure $c_{44}$ and $c_{12}$ are the ductile and brittle surrogates, respectively, and *D*-parameter uses $\gamma_{usf}$ and $\gamma_{surf}$, respectively, for ductility and brittle behavior. Even VEC, while used on its own as a ductility metric, is similar in that shear modulus generally increases between Group IV, V, and VI elements, i.e., $G_{IV} < G_V < G_{VI}$. Generally speaking, ductility is lower in BCC R-CCAs with larger fractions of Groups V and VI elements. It is therefore worthwhile to examine how well ductility correlates with various material properties. Figure 4 presents the correlation coefficients between compressive and tensile strain to failure and several elastic and thermodynamic properties, calculated via a molar RoM. The *D*-parameter is also shown in Figure 4 for comparison. Nearly all material properties show good (inverse) correlation to compressive plasticity, with most properties exceeding $|R_P| > 0.7$, comparable to that of *D*, $R_P = +0.71$. In tension, the *D*-parameter ($R_P = +0.63$) is superior to all the RoM materials properties, although $\Delta H_v$, and $\Delta H_a$ have comparable correlation with tensile failure strain, each with $R_P$ near -0.6. Interestingly, Poisson's ratio, $\nu$, exhibits a very strong positive correlation with compressive plasticity, $R_P = +0.86$, but a very weak negative correlation in tension, $R_P = -0.17$. It is not surprising that the correlation of any parameter with tensile ductility is lower than compressive plasticity due to extrinsic effects such as grain boundary embrittlement, to be discussed later. However, the clear change in sign exhibited by $\nu$ is highly unexpected. This trend may be explained by the compositions of the alloys that have been investigated in tension, primarily soft, Senkov-derivative alloys with high amounts of Hf, Ti and Zr. Since these elements are HCP under standard conditions and often reported as mechanically unstable in the BCC phase, the elastic moduli for the HCP phase, used in the RoM calculations, may lead to erroneous values for elastic properties, specifically ν. However, the strong correlation of ν with compressive plasticity in Group V and VI rich alloys may indicate that a more detailed investigation of the correlation between tensile failure strain and DFT-calculated ν is warranted.

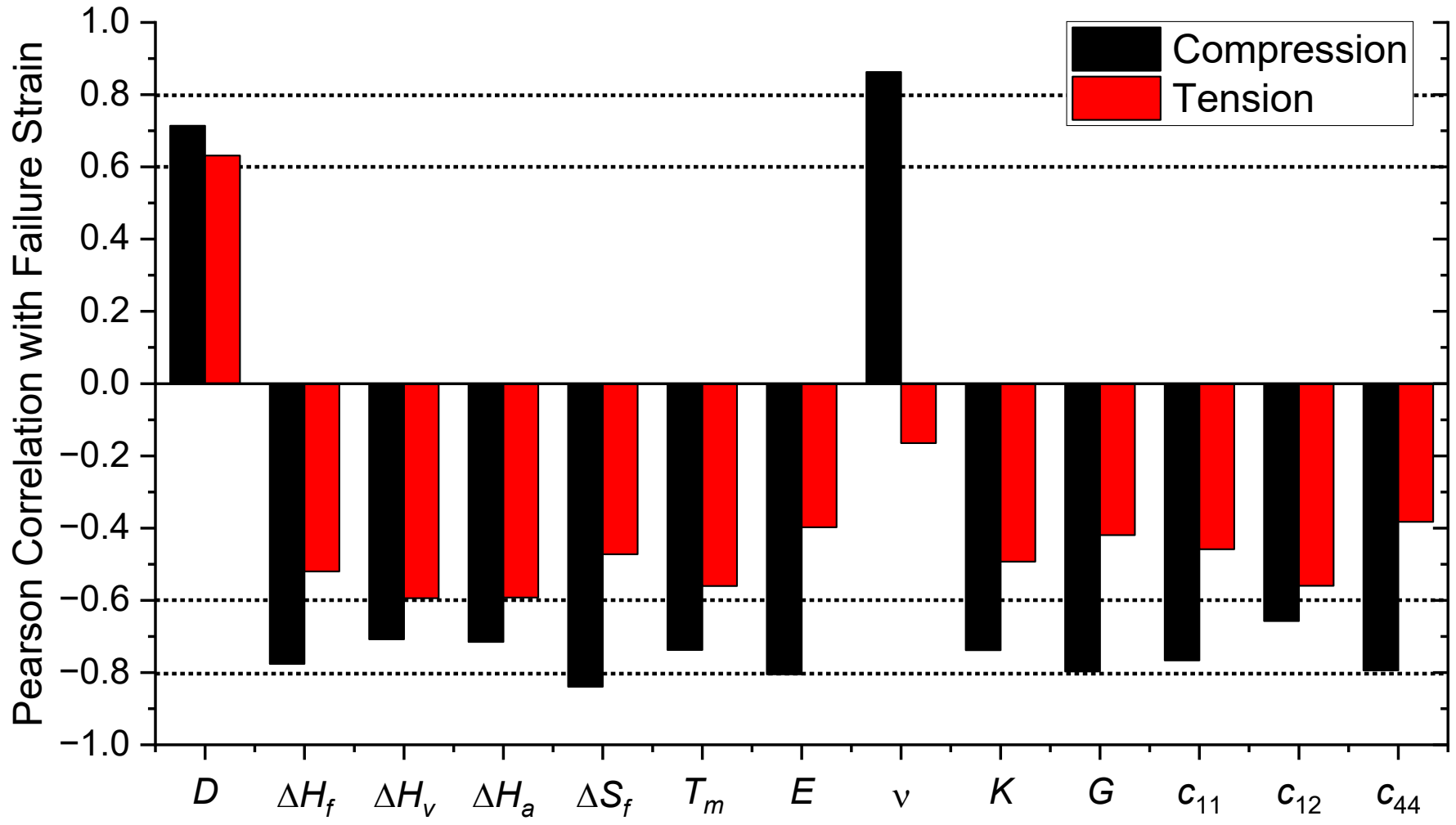

*Figure 4: Pearson correlation coefficients between compressive and tensile failure strain and thermodynamic and elastic properties of the R-CCA alloys. The dashed lines are meant as a guide across the parameters.*

Figure 5 presents the correlation between compressive and tensile plasticity and several "atomic" level properties calculated using a molar RoM, specifically, average atomic diameter ($d_{avg}$) lattice distortion (δ), Allen, Pauling, and Allred-Rochow electronegativity, the "electronegativity distortion" ($\chi_{\delta,EN}$, where *EN* indicates the particular electronegativity scale), i.e., a measure of the variation in electronegativity in the alloy calculated by replacing $d_i$ and $\bar{d}$ in Eq. (1) with $\chi_i$ and $\bar{\chi}$, and the electronegativity corrected average atomic diameter and lattice distortion ($d_{avg,EN}$ and $\delta_{\chi,EN}$, respectively). VEC is also presented for comparison. Consistent with all properties presented above, correlation with most atomic level properties is better with compressive failure strain than in tension. Pauling electronegativity and the associated $\delta_{\chi,\mathrm{Pauling}}$ and $\chi_{\delta,\mathrm{Pauling}}$ all show a strong inverse correlation in compression and weak correlation in tension, while the correlations of Allen and Allred-Rochow parameters are mixed. Allred-Rochow electronegativity and $\delta_{\chi,\mathrm{Allred}}$ exhibit very weak to weak inverse correlation in both tension and compression. Allen negativity and its derived parameters all have weak, but clear correlation with tensile ductility, with $|R_P|$ of ~0.4, and much weaker, if any, correlation with compressive plasticity.

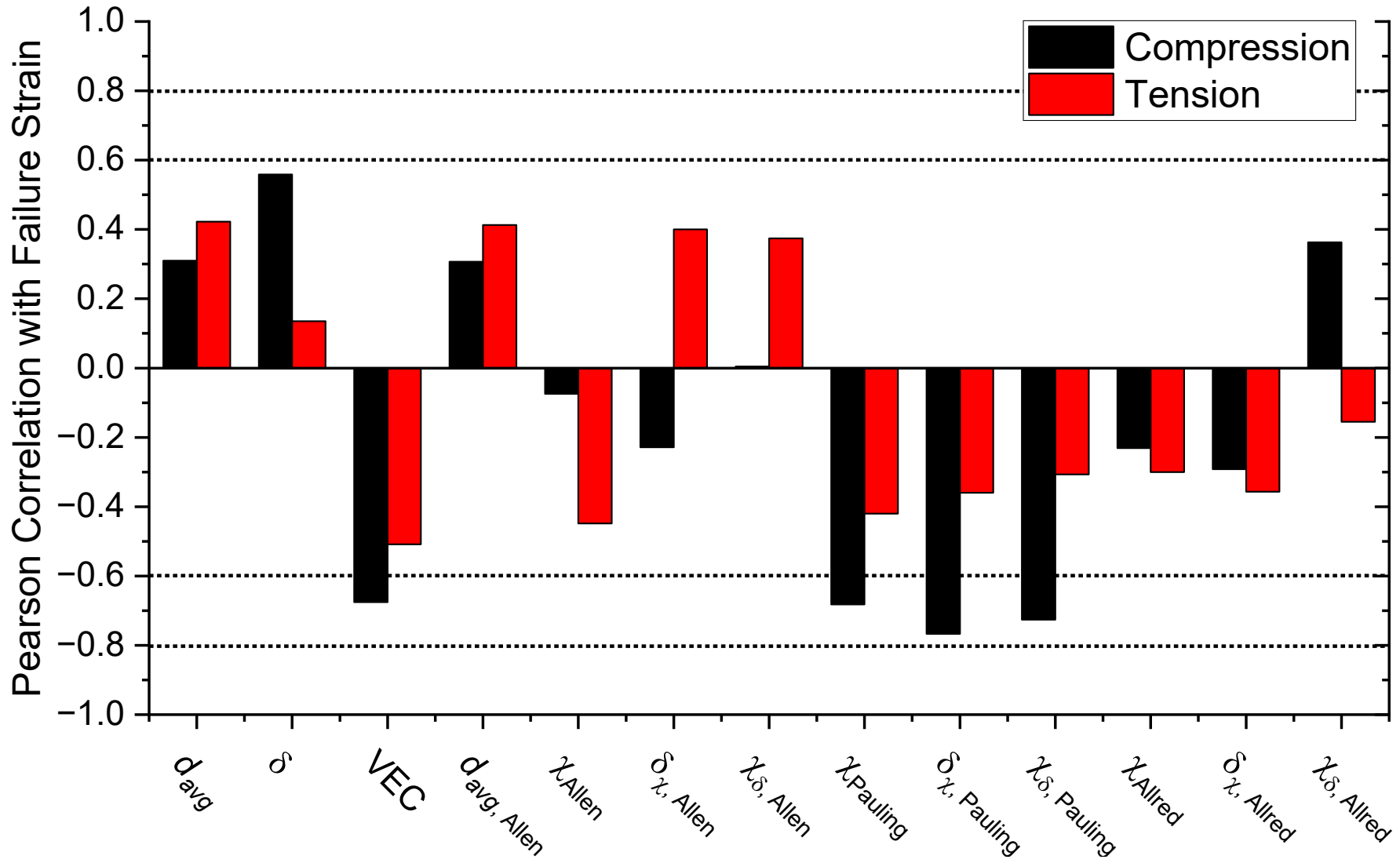

*Figure 5: Pearson correlation coefficients between compressive and tensile failure strain atomic diameter, lattice misfit, electronegativity, electronegativity misfit, and electronegativity-corrected lattice misfit of R-CCA alloys. The dashed lines are meant as a guide across the parameters.*

### 3.3. $\Delta G_{BCC \to HCP}$ Correlations with Ductility

Figure 6 plots compressive and tensile strain to failure versus $\Delta G_{B \to H}$, calculated using both the Thermo-Calc TCHEA8 and Pandat PanRHEA2024 databases. Both databases yield good correlation with both compression and tension, comparable to that of VEC but not as strong as *D*. TCHEA8 performs slightly better in tension, while PanRHEA2024 performs better in compression ($\Delta G_{B \to H}$ calculated using TCHEA8 are used in the remainder of the current work). Still, the correlation of the *D*-parameter is better than $\Delta G_{B \to H}$ calculated using either database.

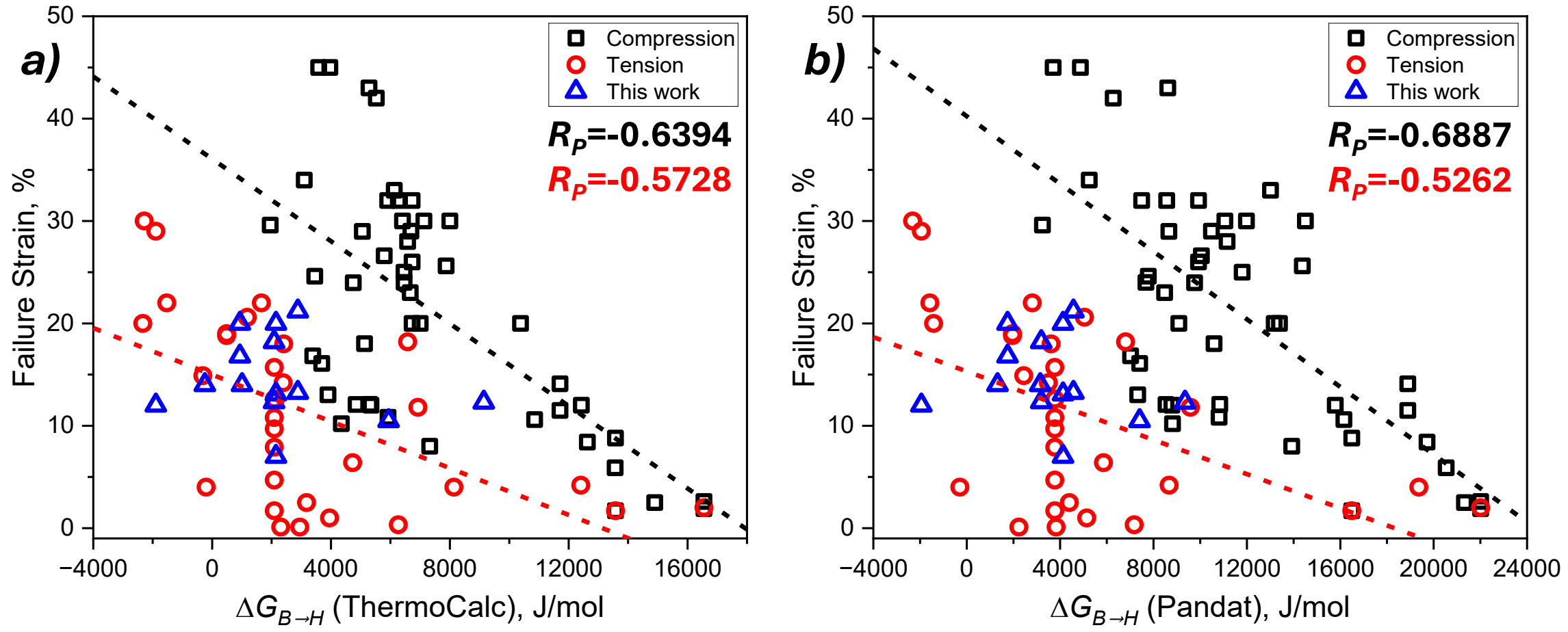

*Figure 6: Plots of compressive (black squares) and tensile (red circles) strain to failure versus $\Delta G_{B \to H}$ calculated using a) the ThermoCalc TCHEA8 database and b) the Pandat PanRHEA database. Blue triangles are experimentally measured tensile ductility of alloys identified in the present work using $\Delta G_{B \to H}$.*

Figure 6 illustrates that there is a relationship between mechanical behavior and $\Delta G_{B \to H}$. While changes in chemical potential represent one contribution to the total energy barrier to dislocation slip, others exist. Strain fields associated with solid solution strengthening, precipitates, or other dislocations also contribute to strengthening and reduced dislocation motion. Lattice distortion, δ, the extension of solid solution strengthening to CCAs, is particularly important in single-phase BCC R-CCAs. Figure 7*a* presents the relationship between ductility and $\delta \cdot \Delta G_{B\rightarrow H}$. Multiplication of $\Delta G_{B\rightarrow H}$ by lattice distortion slightly improves correlation with tensile failure strain, but the correlation with compressive failure strain is significantly reduced. However, if the effects of (Allen) electronegativity differences between constituent elements are considered, and $\Delta G_{B\rightarrow H}$ is instead multiplied by $\delta_\chi$, the correlation with compressive failure strain becomes excellent, $R_P$ = -0.82; while in $R_P$ = -0.66 in tension, a stronger correlation than any other parameters, including *D*.

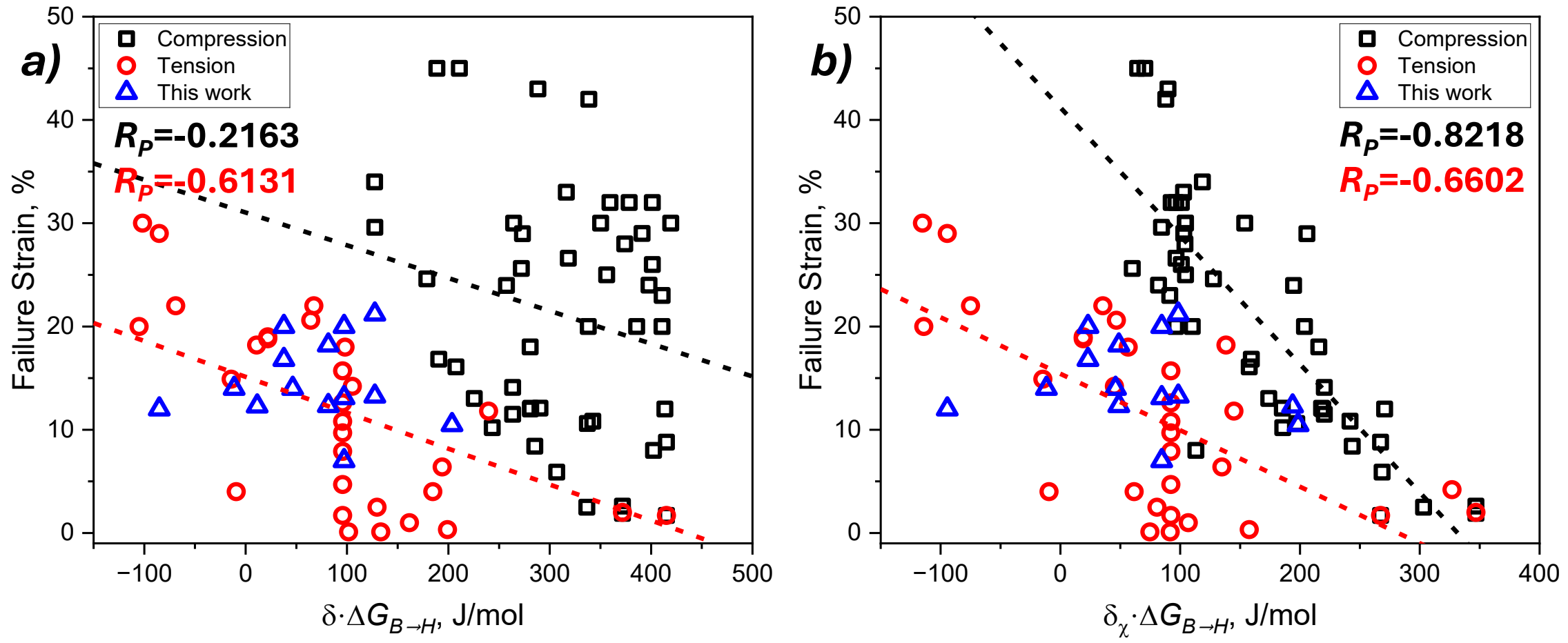

*Figure 7: Plots of compressive (black squares) and tensile (red circles) strain to failure versus a) $\delta \cdot \Delta G_{B\rightarrow H}$ and b) $\delta_\chi \cdot \Delta G_{B\rightarrow H}$ ($\Delta G_{B\rightarrow H}$ calculated using TCHEA8). Blue triangles are experimentally measured tensile ductility of alloys identified in the present work using $\Delta G_{B\rightarrow H}$.*

If $\delta_\chi \cdot \Delta G_{B\rightarrow H}$ is considered a proxy for the ability for a material to plastically deform, and a surrogate for brittle is introduce, $\gamma_{surf}$ as defined for the Rice parameter above, a non-dimensional ductility parameter is defined, $\Gamma = (\delta_\chi \cdot \Delta G_{B\rightarrow H})/\gamma_{surf}$. Figure 8 presents the relationship between Γ and tensile and compressive strain to failure. The correlation between Γ and compressive plasticity is slightly reduced compared to $\delta_\chi \cdot \Delta G_{B\rightarrow H}$, -0.80 compared to -0.82, but the correlation with tensile ductility is improved from -0.66 to -0.70.

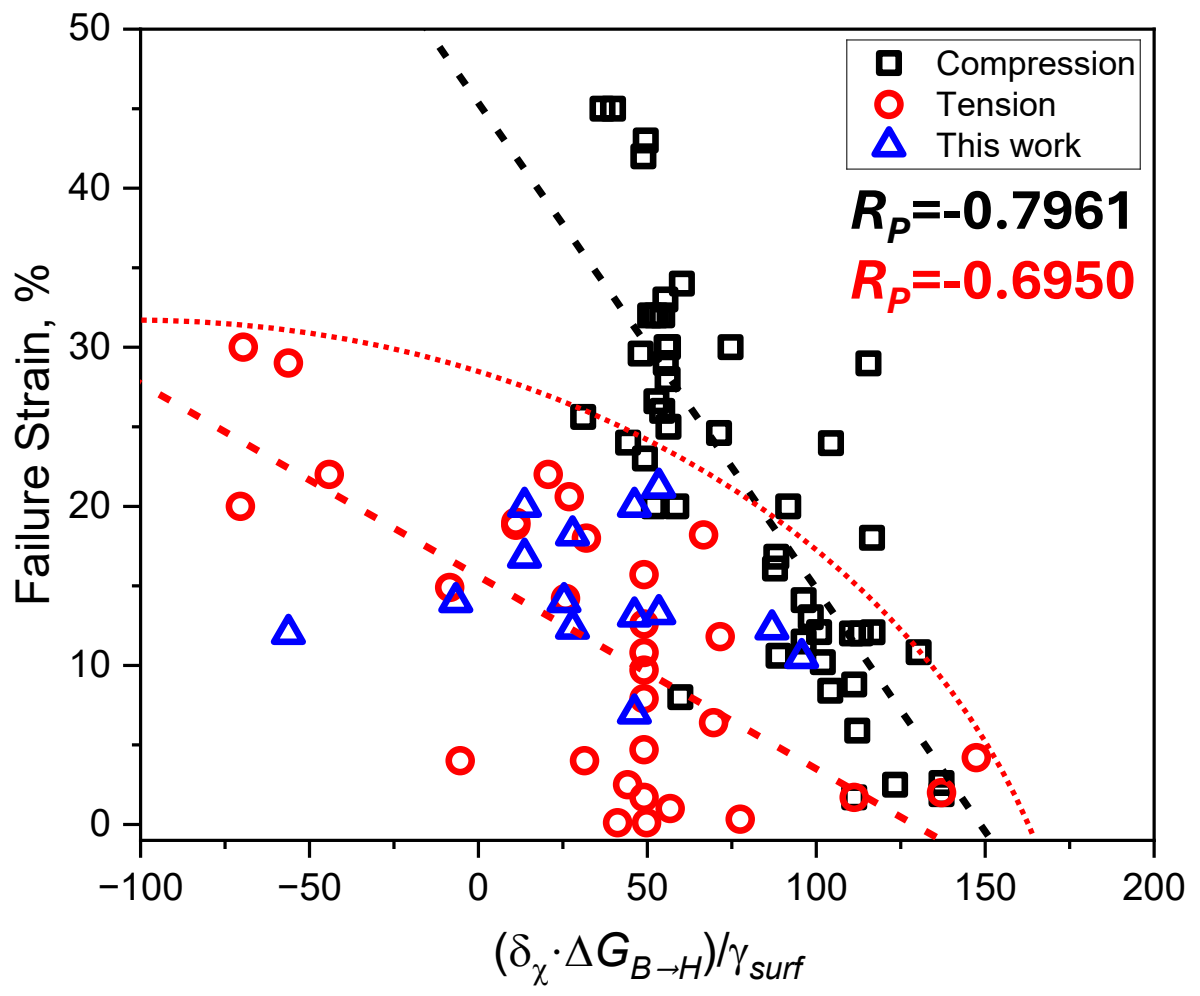

*Figure 8: Plots of compressive (black squares) and tensile (red circles) strain to failure versus $\Gamma = (\delta_\chi \cdot \Delta G_{B\rightarrow H})/\gamma_{surf}$ ($\Delta G_{B\rightarrow H}$ calculated using TCHEA8). Blue triangles are experimentally measured tensile ductility of alloys identified in the present work using $\Delta G_{B\rightarrow H}$. The dotted line is drawn as an approximation of "possible" tensile ductility if extrinsic effects are eliminated.*

## 4. Discussion

Several insights can be made through the analysis of the correlation of ductility metrics and material properties with tensile and compressive strain to failure. Most notably, except for Cauchy pressure, all previously reported ductility metrics exhibit excellent correlation with compressive plasticity, but the correlation with tensile ductility is weaker. This is not unexpected since the scatter in experimental data is greater in tension due to extrinsic factors affecting macroscopic ductility, in particular grain boundary embrittlement. However, the decrease is most notable for Pugh ratio and Rice parameter, dropping from $|R_P| > 0.85$ to 0.15 and 0.26, respectively. This is most likely because these two parameters are dependent on a RoM calculation of $G$ and $K$ using the elastic moduli for Hf, Ti, and Zr in the HCP crystal structure.

The correlation between Cauchy pressure (calculated using the HCP single crystal elastic constants of Hf, Ti, and Zr) and failure strain is puzzling, not only because the correlation is better in tension, but also because Cauchy pressure was recently reported to be essentially equivalent to Pugh ratio [68]. Senkov and Miracle found a linear relationship between Pugh ratio and Cauchy pressure divided by $K$. Indeed, the correlation between $(c_{12} - c_{44})/K$ and the Pugh ratio are 0.98 and 0.79, respectively, for the compression and tension data (see SM Table S19). If the single crystal elastic constants are used to calculate $G_{Reuss}$ (or $G_{Voigt}$) and $K_{c_{ij}} = (c_{11} + 2c_{12})/3$ for the Pugh ratio, Pugh$_2$ = $G_{Reuss}/K_{c_{ij}}$, both Pugh$_2$ and $(c_{12} - c_{44})/K_{c_{ij}}$ are derived from the same starting elastic constants ($c_{11}$, $c_{12}$, and $c_{44}$). The correlation between Pugh$_2$ and $(c_{12} - c_{44})/K_{c_{ij}}$ are 1.00 and 0.99 for compression and tensile datasets, respectively (SM Table S22). The quantity $(c_{12} - c_{44})/K_{c_{ij}}$ also correlates well with compressive failure stain (0.89) and poorly with tensile ductility (-0.22), just like Pugh ratio, +0.89 and +0.16, respectively in compression and tension. It is quite puzzling how $(c_{12} - c_{44})$ can exhibit no correlation with compressive plasticity and good correlation in tension, but when divided by $K$, $(c_{12} - c_{44})/K_{c_{ij}}$ can show outstanding correlation in compression but poor correlation with tensile ductility. One conclusion can be made from the analysis presented in the present work: a RoM estimate from the elastic properties of pure components should not be used if Group IV elements are considered. This may be one reason for the inability of previously reported ductility metrics to accurately estimate the ductility of R-CCAs. This is especially true for tensile ductility, since most reported tensile experiments have been performed on R-CCAs containing significant fractions of Hf, Ti, and/or Zr.

One question that has arisen during the course of this work is "Why does Allen electronegativity, $\chi_{Allen}$, and $\delta_{\chi,Allen}$ yield the best correlation between failure strain and Γ-metric?" Both $\delta_{\chi,Pauling}$ and $\delta_{\chi,Allred}$ were considered (SM Table S23), and Pauling electronegativity, $\chi_{Pauling}$, and $\delta_{\chi,Pauling}$ exhibit much stronger correlations with both compressive and tensile strain to failure. However, $\chi_{Pauling}$ was developed based on covalent bonding, or electron transfer between atoms making a bond. Therefore, it may not be appropriate for describing the largely metallic nature of bonding in BCC R-CCAs. Instead, $\chi_{Allen}$ is much simpler, based on the average energy of the valence electrons for a free atom of a given species. Other electronegativities provide different measures of the relative bond strength between electron and nucleus: *i*) Allred-Rochow is based on the effective nuclear charge experienced by valence electrons; *ii*) Mulliken-Jaffe (and Sanderson) is based on the average energy to remove or add an electron, ionization energy and electron affinity, respectively. Still, the "average valence electron energy" of the Allen electronegativity seems to provide the most appropriate estimate of the tendency of elements to share electrons in metallic bonding.

Another general conclusion that can be made from the above analysis is that most bulk thermodynamic or elastic properties exhibit a strong correlation with compressive failure strain when considered by itself, Figure 4. The correlations of these RoM properties with tensile ductility are also comparable or better than Pugh ratio, Cauchy pressure, Rice parameters, and even VEC and HCP fraction. It is clear that getting a "80 % solution" to the question "Will an R-CCA be ductile?" can readily be achieved. Achieving any better predictive capability appears to be much more difficult, even when DFT is utilized, with $|R_P|$ for $D$ being only 0.71 and 0.63 for compressive and tensile plasticity, respectively. The proposed Γ-metric that utilizes a CALPHAD thermodynamic basis plus a RoM estimation of electronic and lattice distortion contributions performs quite well, slightly better than $D$. However, the improvement is modest, $|R_P|$ of 0.80 and 0.70 for compression and tension, respectively, and the statistical significance of the improved correlation is unclear. What should be clear is that utilizing a thermodynamic foundation such as the DFT calculation of unstable stacking fault energy or estimation using CALPHAD or other thermodynamic modeling will provide a more informed estimation of intrinsically ductile behavior than previously proposed ductility metrics.

One thing that all the ductility metrics and material properties discussed herein have in common is that they are average properties of the bulk material. Mechanical behavior, specifically dislocation motion, in R-CCAs may be dictated by the local environment through which they must move. The potential energy landscape experienced by a dislocation in an R-CCA is no doubt very rugged, and the rate controlling factor may be the potential barrier experienced near a certain atomic species, e.g., Mo or W, or near a cluster of atoms (i.e., short-range order). While establishing the exact energy landscape over which a dislocation must traverse is challenging, consideration of the largest energy barriers for dislocation motion, their concentration, and their average separation may provide insight into dislocation mobility in R-CCAs. For instance, Group VI elements are known to be the most detrimental to intrinsic ductility. This is due in large part to the increased number of $d$-band electrons they possess, compared to Group IV and V elements, which give rise to stronger, stiffer, directional bonds and much higher Peierls stresses [24,41]. Quantifying the potency of these elements, in terms of lattice distortion, modulus mismatch, changes in the local electron density and bond directionality, and their concentration within the lattice may provide more information on dislocation mobility and macroscopic intrinsic ductility than bulk average properties.

A comparison of the two equiatomic alloys MoNbTi and NbTaTi serves as an example of the above observation. The Γ-metrics of these two alloys are 31.4 and 66.7, respectively (listed in Table 1), suggesting MoNbTi should possess the greatest ductility. However, NbTaTi exhibits a tensile failure strain 4.5 times larger than MoNbTi, 18 % compared to 4 %. On the other hand, the $D$-parameter can capture the difference between these two alloys, 2.68 for MoNbTi compared to 3.36 for NbTaTi. The difference between the $D$-parameters of these two alloys is largely in $\gamma_{usf}$, 0.82 J/m$^2$ and 0.61 J/m$^2$ for MoNbTi and NbTaTi, respectively, while the surface energies are similar, 2.07 J/m$^2$ and 2.19 J/m$^2$, respectively. The difference in $\gamma_{usf}$ of these two alloys is most likely a result of the rigid, more directional bonds surrounding Mo atoms compared to Ta. The CALPHAD-derived Γ does not appear to capture the increased stiffness and directionality of the Group VI element. This may be explained in part by the calculated $\Delta G_{B\to H}$ (using the TCHEA8 database) for the two alloys, 8137 J/mol and 6582 J/mol for MoNbTi and NbTaTi, respectively. $\Delta G_{B\to H}$ is larger for MoNbTi, suggesting dislocation motion should be more difficult, but the difference is smaller than expected for Group VI containing alloys, e.g., 22015 J/mol for MoNbTaW. Indeed, using the PanRHEA2024 database $\Delta G_{B\to H}$ for MoNbTi is considerably larger than using TCHEA8, 19366 J/mol, compared to 6805 J/mol for NbTaTi (close to the TCHEA8).

Using the PanRHEA2024 $\Delta G_{B\rightarrow H}$ yield Γ of 74.7 and 68.9 for MoNbTi and NbTaTi, respectively, correctly describing the observed difference in tensile failure strain between the two alloys. Therefore, uncertainty in the assessed CALPHAD thermodynamic functions may account for discrepancy between Γ and tensile failure strain of these two alloys. Another possible reason for the discrepancy between Γ and tensile ductility of MoNbTi and NbTaTi is $\delta_\chi$, which is expected to decrease from 2.10 % to 0.76 % in MoNbTi but increase from 0.17 % to 2.27 % in NbTaTi. NbTaTi is known as a low lattice distortion alloy, so it's large increase in $\delta_\chi$ is a possible shortcoming of using the RoM electronegativity-corrected lattice distortion. A better estimate for this parameter may be useful for future corrections. In summary, the DFT calculations used in the determination of $\gamma_{usf}$ in the *D*-parameter likely captures the physics of bond characteristics and local atomic structure better than $\Delta G_{B\rightarrow H}$ or any RoM property. Still, the simplicity of using $\Delta G_{B\rightarrow H}$ and Γ remain attractive.

*Table 1: Tensile failure strain and calculated ductility metrics and associate properties for equiatomic NbTaTi and MoNbTi.*

| Alloy | NbTaTi | MoNbTi |
|---|---|---|
| Tensile Failure Strain | 18.2 % | 4.0 % |
| *D*-parameter | 3.36 | 2.68 |
| Γ-metric, TCHEA8 | 66.7 | 31.4 |
| $\Delta G_{B\rightarrow H}$, TCHEA8 (J/mol) | +6582 | +8137 |
| Γ-metric, PanRHEA | 68.9 | 74.7 |
| $\Delta G_{B\rightarrow H}$, PanRHEA (J/mol) | +6805 | +19366 |
| $\delta$ | 0.17 % | 2.10 % |
| $\delta_\chi$ | 2.27 % | 0.76 % |

One thing that has been ignored throughout the above analysis is extrinsic ductility. In particular, interstitial solutes and grain boundaries can dramatically affect mechanical behavior [4]. Oxygen and nitrogen additions have been found to enhance mechanical properties, including ductility, when added to soft, Senkov-derivative alloys [69–71]. However, their presence even in small amounts can cause brittle, intergranular failure in Group V and VI elements and their alloys [4]. This raises the question: *Is intrinsic ductility a moot point in R-CCAs because extrinsic effects will always be the limiting factor for useable ductility in R-CCAs that exhibit high-temperature strength retention?* Even if intrinsic ductility is of some importance, extrinsic factors must still be considered. Recent updates to CALPHAD-based thermodynamic databases have introduced oxygen (and nitrogen), which allows for O-solubility to be calculated, which could be used as a first order approximation of the propensity for an alloy to exhibit grain boundary embrittlement due to O segregation to grain boundaries. Figure 9*a* presents the compressive and tensile failure strain of R-CCAs reported in literature as a function of the O-solubility calculated at 1200 °C. While the correlation between failure strain and O-solubility is not as strong as for *D*-parameter and Γ-metric there is a clear relationship: alloys with a higher O-solubility tend to exhibit greater strain to failure. If O-solubility is combined with Γ, Figure 9*b*, it can be seen that alloys with reasonable room temperature ductility (>5 % in tension or >25 % in compression) are bound by approximately $\Gamma < 100$ and $w_O^{1200} < 10^{-7}$ wt. % for compression and $\Gamma < 100$ and $w_O^{1200} < 10^{-5}$ wt. % for tension. While not a rigorous treatment of the effect of oxygen, it may provide a first order guideline for estimating extrinsic tensile brittleness.

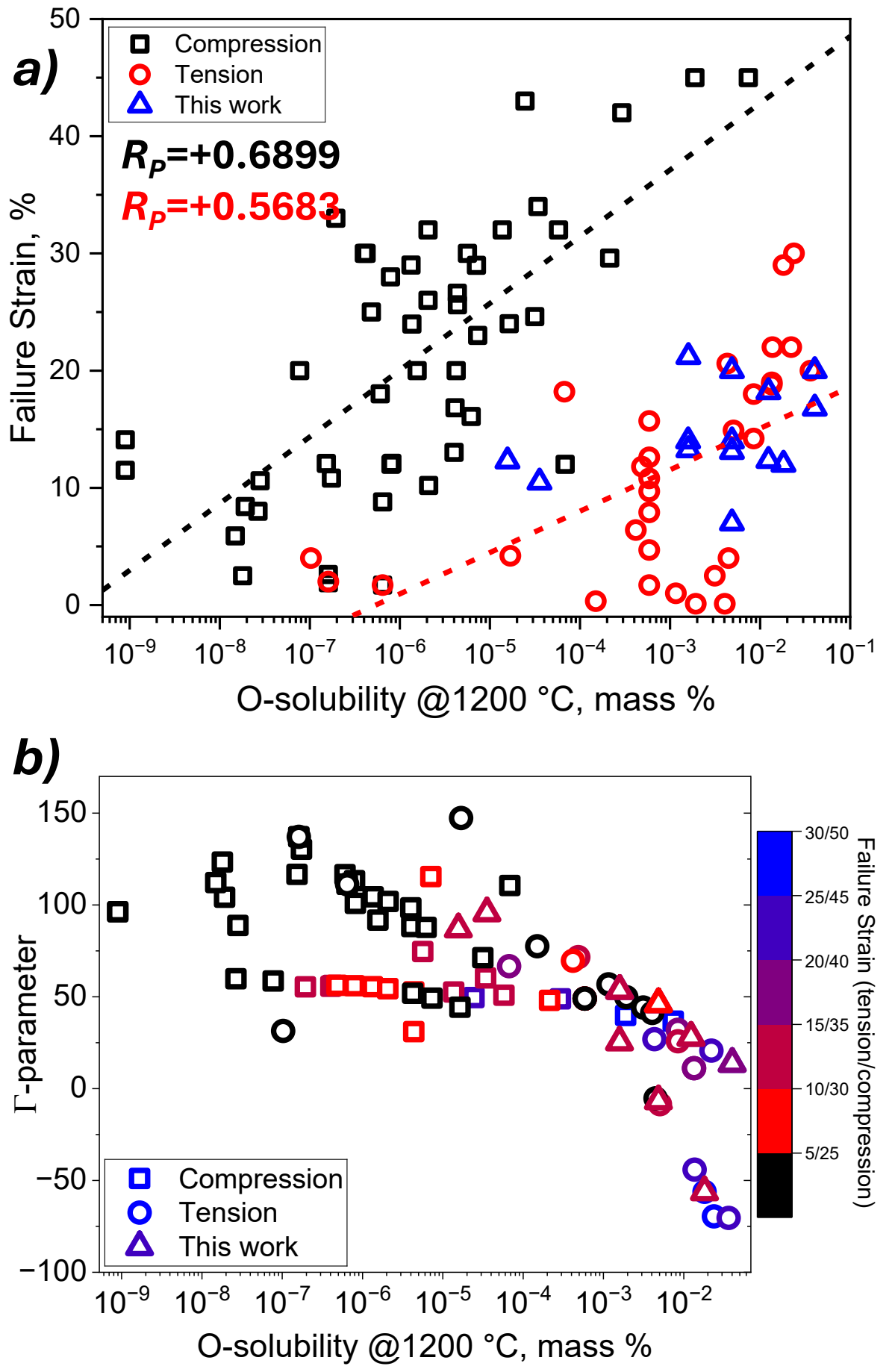

*Figure 9: a) Compressive (black squares) and tensile (red circles) strain to failure versus oxygen solubility at 1200 °C (calculated using TCHEA8). Blue triangles are experimentally measured tensile ductility of alloys identified in the present work using ($\Delta G_{B\rightarrow H}$. b) The oxygen solubility at 1200 °C and Γ-parameter for R-CCAs with reported compressive (squares) and tensile (circles) strains to failure, and of the R-CCAs investigated in the present study (triangles). The color of the symbol indicates the reported failure strain.*

Finally, it should also be mentioned that because of extrinsic effects such as grain boundary embrittlement, experimentally measured tensile ductility can vary substantially. Therefore, the tensile ductility measurements presented in Figure 8 should be thought of as "potential" ductility. That is, the largest experimentally observed failure strain for a particular alloy (and Γ) indicates the maximum possible ductility, if the alloy is perfectly fabricated and thermomechanically processed, without the introduction of extrinsic embrittling factors such as O contamination. An estimate of this "possible" ductility is indicated by the dotted red line in Figure 8.

## 5. Conclusions

This work examined the thermodynamic and physical origins of room-temperature ductility in BCC refractory compositionally complex alloys by comparing reported and newly measured tensile and compressive failure strains with established ductility criteria, RoM properties, CALPHAD phase stabilities, and descriptors of local lattice and electronic disorder. The principal conclusions are summarized below.

- Most established ductility metrics correlate substantially better with compressive strain to failure than with tensile strain to failure. The Pugh and Rice criteria are especially sensitive to this distinction: both exhibit strong correlations in compression but weak correlations in tension. Their reliance on RoM elastic properties is problematic for alloys containing Hf, Ti, and Zr because reliable elastic constants for the metastable BCC forms of these elements are generally unavailable and HCP values are commonly substituted. Cauchy pressure, as conventionally applied, shows essentially no correlation with compressive plasticity and is not a reliable standalone criterion for R-CCAs.
- The Gibbs free-energy difference between BCC and HCP, $\Delta G_{B\to H}$, provides a physically motivated connection between phase stability and dislocation mobility. BCC slip can sample local configurations related to HCP and other displacive products; consequently, $\Delta G_{B\to H}$ serves as a proxy for the energetic landscape encountered during deformation. Values calculated independently with the TCHEA8 and PanRHEA2024 databases correlate with both compressive and tensile failure strain, supporting the general thermodynamic premise despite database-dependent differences.
- Accounting for local chemical and size disorder improves this thermodynamic description. Multiplying $\Delta G_{B\to H}$ by the electronegativity-corrected lattice distortion improves correlation with both tensile and compressive plasticity, while normalization by surface energy produces the nondimensional parameter $\Gamma = \left(\delta_\chi \cdot \Delta G_{B\to H}\right)/\gamma_{surf}$, with correlations of −0.80 and −0.70, respectively, in compression and tension.
- The improvement over the *D*-parameter is modest and its statistical significance requires confirmation with larger, consistently processed datasets; nevertheless, Γ provides a readily accessible framework that combines resistance to dislocation-mediated plasticity with resistance to fracture.

- Intrinsic ductility metrics alone cannot determine measured tensile ductility. Grain boundaries, segregation, processing history, and interstitial contamination can suppress tensile strain to failure well below the intrinsic or "potential" ductility indicated by Γ. Calculated oxygen solubility offers a useful first-order screen for this extrinsic limitation: alloys with greater oxygen solubility tend to sustain greater strain to failure, whereas intrinsically promising alloys with low oxygen tolerance remain vulnerable to grain-boundary embrittlement. Effective R-CCA design should therefore optimize both the intrinsic thermodynamic deformation landscape and tolerance to extrinsic embrittling species.

**Acknowledgements**

This work was funded primarily by the Army Research Office through the DoD Defense Established Program to Stimulate Competitive Research program (AFOSR-FA9550- 23-1-0503). EAL and CJCN also acknowledge support by the National Science Foundation Materials Research Science and Engineering Center program through the UT Knoxville Center for Advanced Materials and Manufacturing (DMR-2309083). CT and YJH acknowledge support by the National Science Foundation under the award DMR-2442768. Lia Amalia and John Whitlow at UTK are also acknowledged for their contributions in fabricating and testing some of the alloys presented in this work.